\documentclass[letter,twocolumn]{jpsj3_cm}
\usepackage{txfonts}
\usepackage{amsmath}%
\usepackage{bm}%

\def\tn{$T_{\rm N}$}
\def\tsn{$T^*_{\rm N}$}
\def\qone{${\bm q}_{1}$}
\def\qop{${\bm q}^*_{1}$}
\def\bb{${\bm B}$}
\def\qq{${\bm Q}$}
\def\degr{$^{\circ}$}
\def\qa{${\bm q}_{A}$}
\def\qoeq{${\bm q}_{1}$=$(0.2, 0.3, 0)$}
\def\qopeq{${\bm q}^*_{1}$=$(0.2, 0.3, {\delta})$}
\def\qaeq{${\bm q}_{A}{\simeq}({\pm}0.09, {\pm}0.20, {\mp}0.28)$}

\title{Unique Helical Magnetic Order and Field-Induced Phase in Trillium Lattice Antiferromagnet EuPtSi}

\author{Koji Kaneko$^{1,2}$\thanks{kaneko.koji@gmail.com}, Matthias D. Frontzek$^3$, Masaaki Matsuda$^3$, Akiko Nakao$^4$, Koji Munakata$^4$,\\ Takashi Ohhara$^2$, Masashi Kakihana$^5$,  
Yoshinori Haga$^6$, Masato Hedo$^7$, Takao Nakama$^7$, and\\ Yoshichika \=Onuki$^7$}
\inst{$^1$Materials Sciences Research Center, Japan Atomic Energy Agency, Tokai, Naka, Ibaraki 319-1195, Japan \\
$^2$J-PARC Center, Japan Atomic Energy Agency, Tokai, Naka, Ibaraki 319-1195, Japan \\
$^3$Neutron Scattering Division, Oak Ridge National Laboratory, Oak Ridge, Tennessee 37831, U. S. A. \\
$^4$Comprehensive Research Organization for Science and Society, Tokai, Ibaraki 319-1106, Japan \\
$^5$Graduate School of Engineering and Science, University of the Ryukyus, Nishihara, Okinawa 903-0213, Japan \\
$^6$Advanced Science Research Center, Japan Atomic Energy Agency, Tokai, Naka, Ibaraki 319-1195, Japan \\
$^7$Faculty of Science, University of the Ryukyus, Nishihara, Okinawa 903-0213, Japan \\
} 

\abst{
Magnetic transition phenomena in cubic chiral antiferromagnet EuPtSi with {\tn}=4.0~K were investigated by means of single crystal neutron diffraction. 
At 0.3~K in the ground state, magnetic peaks emerge at positions represented by an ordering vector {\qoeq} and its cyclic permutation. 
Upon heating, an additional magnetic peak splitting with hysteresis was uncovered at around $T^*_{\rm N}{\sim}$2.5~K, 
indicating the presence of a first-order commensurate-incommensurate transition with {\qopeq} (${\delta}_{\rm max}{\simeq}$0.04) at {\tsn}.
A half-polarized neutron scattering experiment for polarization parallel to the scattering vector revealed that polarization antiparallel to the scattering vector has stronger intensity in both magnetic phases. 
This feature clarifies the single chiral character of the helical structure with moments lying perpendicular to the ordering vector in both ordered states.
Under a vertical magnetic field of 1.2~T for {\bb}${\parallel}$[1,1,1] at 1.9~K entering into the so-called $A$ phase, magnetic peaks form characteristic hexagonal patterns in the equatorial scattering plane around nuclear peaks. 
An ordering vector {\qaeq} of the $A$-phase has similar periodic length as {\qone}, and could be the hallmark of a formation of skyrmion lattice in EuPtSi. 
}

\begin{document}
\maketitle

The lack of spatial inversion symmetry in a structure may lead to emergence of unconventional states of matter.
Non-centrosymmetry results in appearance of antisymmetric interaction, such as the Dzyaloshinsky$-$Moriya (DM) interaction\cite{Dzyaloshinsky1958,Moriya1960}.
The DM interaction often stabilizes nontrivial magnetic structures.
In multiferroic compounds, this interaction give rise to an intimate coupling between spiral magnetic structures and ferroelectric polarization\cite{Katsura2005,Arima2011}.
In cubic $B20$-type compounds crystallizing in the space group $P2_13$, a combination of the DM and ferromagnetic interactions results in a helimagnetic structure with long pitch, as can be seen in the prototypical compound MnSi.\cite{Ishikawa1976,Shirane1983,Ishikawa1984,Ishida1985}
A notable feature of MnSi appears under magnetic fields;
An application of a magnetic field induces a field-induced ordered state with magnetic skyrmions, 
particle-like topologically non-trivial spin textures\cite{Muhlbauer2009}.
This discovery has triggered intensive research on skyrmions, 
and several compounds with the same space group, such as Fe$_{1-x}$Co$_x$Si\cite{Munzer2010,Yu2010}, FeGe\cite{Uchida2008,Wilhelm2011}, Cu$_2$OSeO$_3$\cite{Seki2012,Adams2012}, were found to host skyrmions.

In the so-called trillium lattice, a subgroup of the $B20$ compound with the same space group $P2_13$, 
the atoms at the $4a$ site form a three-dimensional network of corner-sharing equilateral triangles, yielding inherent geometric frustration.
The theoretical study of the classical Heisenberg model with the nearest-neighbor antiferromagnetic interaction in the trillium lattice suggests that 
the ground state lies on the border between an ordered state and a frustrated degenerated state\cite{Hopkinson2006,Isakov2008}.
In the ordered case, the classical Monte Carlo simulation suggests strong first-order character of the transition with prominent partial-order fluctuation above the transition,
and the ordered state could have the characteristic 120$^{\circ}$ structure.
In addition, another study that assumed a ferromagnetic Ising model proposed a new type of spin-ice structure\cite{Redpath2010}.
These outcomes attract interest toward the ground state of trillium lattice in actual compounds. 

EuPtSi is reported to have the cubic LaIrSi-type structure with the space group $P2_13$\cite{Adroja1990}. 
All atoms occupy the $4a$ site, that is, EuPtSi can be classified as the trillium lattice compound.
In contrast to MnSi with dominant ferromagnetic interaction, an antiferromagnetic order is reported in EuPtSi\cite{Kakihana2017,Franco2017,Kakihana2018}.
M\"ossbauer spectroscopy and magnetization measurements indicate a stable divalent state of Eu ions with a large spin moment of $S=7/2$.
While the slightly positive Curie$-$Weiss temperatures suggest ferromagnetic correlation, 
an antiferromagnetic order is identified at relatively low temperatures, {\tn}=4.1~K in EuPtSi.
The first-order-like nature of the transition at {\tn} is corroborated by a sharp peak in the specific heat and steep drops in the electrical resistivity and the magnetic susceptibility.
In addition, the magnetic entropy released at {\tn} is close to half of $R$ln~8, as expected for Eu$^{2+}$.
These features suggest the presence of strong fluctuation above {\tn} leading to the first-order nature of the transition, which stems from underlying geometrical frustration.

\begin{figure*}[!ttttt]
\begin{center}
\includegraphics[width=16cm]{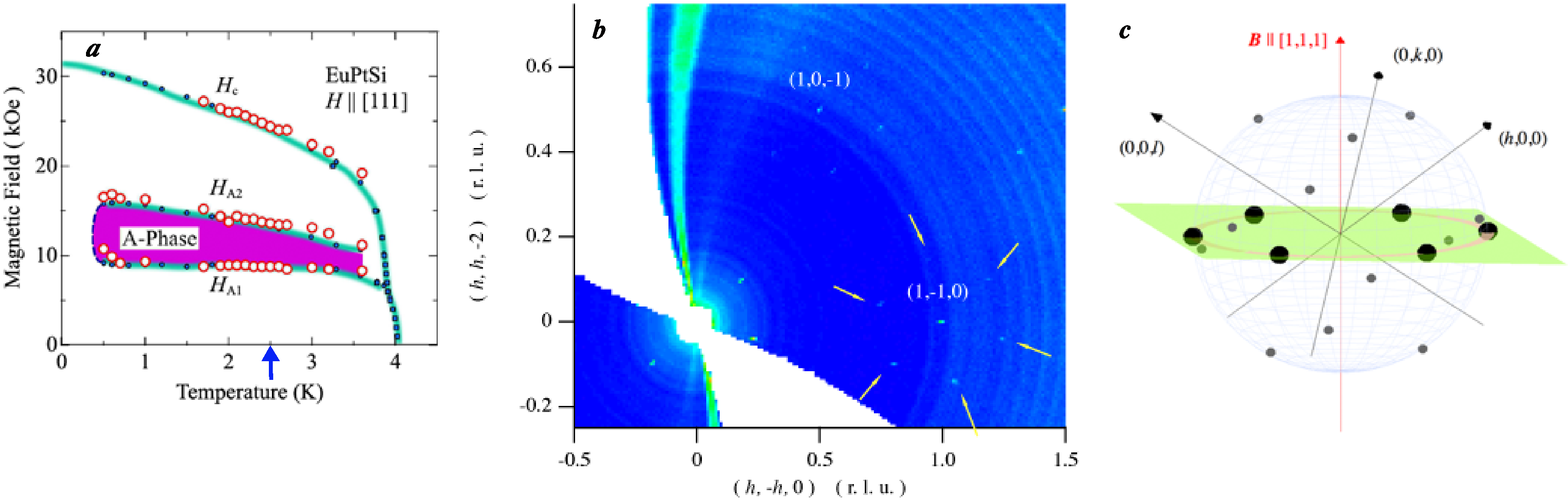}
\caption{(Color online) (a) The $B$-$T$ magnetic phase diagram of EuPtSi for ${\bf B}{\parallel}$[1,1,1] by Kakihana \textit{et al}\cite{Kakihana2018}.  The arrow indicates the additional transition revealed in this study.
(b) A part of the neutron scattering intensity map in the horizontal scattering plane defined by two orthogonal axes [1,~${\bar 1}$,~0] and [1,~1,~${\bar 2}$] in the $A$-phase at 1.2~T, 1.9~K. 
Yellow arrows indicate the magnetic peaks around  (1,~${\bar 1}$,~0) in the $A$ phase. Vertical strong intensity region across the origin corresponds to the dark angle of the magnet. 
(c) Schematic magnetic peak distribution in a three-dimensional reciprocal lattice space around (0,~0,~0) for the ground state with {\qone} (small circles)  and in the $A$ phase with {\qa} (large circles).  
Green sheet represents the horizontal scattering plane against vertical magnetic field along [1,~1,~1]. Sphere corresponds to the isosurface for $|Q|$=0.34~{\AA}$^{-1}$.\vspace{-1cm}}
\label{fig:phA}
\end{center}
\end{figure*}

Furthermore, bulk measurements using single crystal reveal the appearance of a characteristic field-induced ordered state in EuPtSi\cite{Kakihana2018}.
The field-induced phase exists only in a limited region at finite temperature and field in the phase diagram, as displayed in Fig.~\ref{fig:phA}(a), which is reminiscent of the $A$ phase in MnSi in which the skyrmion lattice is formed. 
Indeed, additional Hall resistivity characteristic to this induced phase was identified, which may arise from topological effect. 
These findings motivate us to study the spin structure of EuPtSi, including the field-induced phase.
In this letter, we report the single crystal neutron diffraction study of EuPtSi. 
The ground state has a helical magnetic structure described by {\qoeq}.
An additional first-order transition to an incommensurate state with {\qopeq} was revealed around {\tsn}${\sim}2.5$~K, whereas no trace of fluctuation was detected above {\tn} in the present study.
By entering into the field-induced ordered $A$ phase, the magnetic peaks shift to the equatorial plane orthogonal to the applied field
and form a characteristic hexagonal pattern.
This feature is similar to MnSi, and supports the possible formation of the skyrmion lattice in EuPtSi.

Single crystals of EuPtSi were grown by the Bridgman method using naturally occurring europium.
The details of sample growth have been published elsewhere\cite{Kakihana2017}.
Several single crystalline samples with typical sizes of ${\sim}2{\times}2{\times}1$~mm$^3$ were used in the following neutron scattering experiments.

Neutron scattering experiments on EuPtSi were performed on several instruments.
Single crystal neutron diffraction measurements at zero field were carried out on two instruments:
the time-of-flight Laue diffractometer SENJU, installed at BL18 of the Materials and Life Science Experimental Facility (MLF) of J-PARC in Tokai, \cite{Ohhara2016}
and the wide-angle neutron diffractometer (WAND) installed at the HB-2C beam port of the High Flux Isotope Reactor (HFIR) at Oak Ridge National Laboratory (ORNL), U. S. A. 
In SENJU, neutrons with a wavelength range from 0.4 to 8~{\AA} were used.
Cylindrical arrays of two-dimensional position-sensitive detectors (PSDs) cover the scattering angle from 12{\degr} to 168{\degr} in the horizontal scattering plane.  
On the other hand, WAND uses a monochromatic beam with a wavelength of 1.48~{\AA} provided by the Ge 3~1~1 monochromator, and has a large one-dimensional PSD with a total coverage of 125{\degr}.
The sample was cooled down below 0.3~K using a $^3$He cryostat.

Additional neutron diffraction experiments under a magnetic field were performed on upgraded WAND$^2$ equipped with a two-dimensional PSD, which covers the scattering angles of 120$^{\circ}$ and ${\pm}7.5^{\circ}$ in the horizontal and vertical directions, respectively\cite{Frontzek2017}.
The sample was inserted into a 5~T vertical field magnet to have fields {\bb}${\parallel}$[1,~1,~1], where the horizontal scattering plane is defined by [$h,~{\bar h},~0$] and [$h,~h,~{\bar 2}$], and cooled down to 1.8~K.
 
In order to obtain insights into the magnetic structures, polarized neutron scattering experiments were performed on the triple-axis spectrometer HB-1, also installed at HFIR at ORNL.
Polarized incident neutrons with energy of $E_i$=13.5~meV were extracted via a Heusler monochromator with the polarization rate of 0.90, 
and the polarization of the scattered beam was not analyzed with the vertically focusing  pyrolytic graphite analyzer.
The spectrometer was fixed at the elastic condition in W configuration with a horizontal collimator sequence of 48'-80'-60'-240', which gives energy resolution of 1.5~meV.
The Helmholtz coil together with the guide field was operated to give neutron polarization parallel or antiparallel to the scattering vector {\qq}.
The Mezei-type spin flipper was placed before the coil, which allowed to flip an incident neutron polarization.

\begin{figure}[t]
\vspace{2cm}
\includegraphics[width=8.8cm]{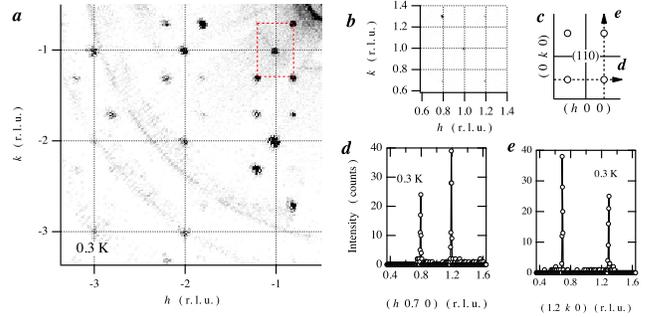}
\vspace{-2cm}
\caption{($a)$ Neutron scattering intensity map of the reciprocal $(h,~k,~0)$ plane of EuPtSi at 0.3~K extracted from the diffraction data recorded on SENJU. 
(b) An up-close view around (1~1~0), and (c) schematic drawing of the observed peak positions, and cross-sectional cut through magnetic peaks for (d) ($h$,~0.7,~0) and (e) (1.2,~$k$,~0) at the same temperature.\vspace{-0.5cm}}
\label{fig:gs}
\end{figure}

A part of the neutron scattering intensity map in the ($h,~k$,~0) scattering plane at 0.3~K measured on SENJU is displayed in Fig.~\ref{fig:gs}(a).
In addition to the nuclear Bragg peaks at integers $h$ and $k$, superlattice peaks were clearly revealed.
Superlattice peaks are prominent in the low $Q$ region, and get weaker with increasing $Q$.
This feature evidences its magnetic origin.
Figure~\ref{fig:gs}(b) displays a magnified map around (1,~1,~0), and the one-dimensional cut along ($h$,~0.7,~0) and (1.2,~$k$,~0) are shown in Fig.~\ref{fig:gs} (d) and (e), respectively.
These data demonstrate that the superlattice peaks were found at (1${\pm}$0.2,~1${\pm}$0.3,~0). 
Specifically, the magnetic ordering vector of EuPtSi in the ground state is {\qoeq}.
Note that no peak was found at (0.7,~0.8,~0). 
In other words, the equivalent ordering vector is limited to a cyclically permutable one, which reflects the symmetry of the crystal structure.
Furthermore, these data prove that the present sample consists of a single domain.

\begin{figure}[!t]
\includegraphics[width=8.8cm]{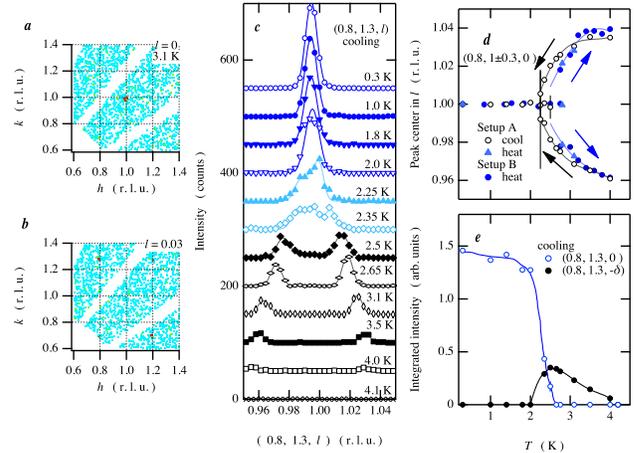}
\caption{(Color online) ($a)$ Neutron scattering intensity maps around (a) (1,~1,~0) and (1,~1,~0.03) measured at 3.1~K. 
(c)  The one-dimensional cut along (0,~0,~$l$) through magnetic peaks at (0.8,~1.3,~0) recorded as a function of temperature upon cooling (bottom to top).
Lines in the figure are the results of three Gaussian fittings.
(d) The obtained peak position $l$ in (0.8,~1.3,~$l$) as a function of temperature measured on SENJU (Setup A, cooling and heating) and WAND (Setup B, only heating process is shown).
 (e) Temperature variation of integrated intensity of (0.8,1.3,~0) and (0.8,1.3,~0) peaks recorded upon cooling process on SENJU.\vspace{-1cm}}
\label{fig:im}
\end{figure}

Upon heating from 0.3~K, additional anomaly was discovered in the neutron scattering data below {\tn}.
The intensity map around (1,~1,~0) at 3.1~K shown in Fig.~\ref{fig:im}(a), which has the same range as Fig.~\ref{fig:gs}(b), reveals the disappearance of magnetic peaks in this ($h,~k,$~0) scattering plane.
Indeed, the magnetic peaks were slightly moved away from the ($h,~k,~$0) plane with a finite $l$ component, as displayed in Fig.~\ref{fig:im}(b), 
where peaks were detected at (1${\pm}0.2, 1{\pm}0.3,{\pm}0.03$); that is, the $h$ and $k$ components remain at the same values. 
In order to unmask the nature of this shift, the detailed temperature variation of the magnetic peak profile was investigated on SENJU and WAND using different single crystals.
Figure~\ref{fig:im}(c) displays the line profile along (0.8,~1.3,~$l$) collected as a function of temperature upon cooling down.
With decreasing temperature, magnetic peaks appear abruptly at 4.0~K around $l{\sim}{\pm}0.04$.
The peak splitting along $l$ becomes smaller with cooling; concomitantly, a central peak at $l$=0 starts to develop below 2.35 K. 
Further decrease in temperature causes the central and satellite peaks to merge into a single peak below 2~K, and remain at the same position down to the base temperature of 0.3~K.
This feature indicates the presence of an incommensurate/commensurate transition at around ${\sim}$2.35~K.
A similar investigation was conducted on the heating process. 
The peak positions in $l$ obtained by three Gaussian fittings at the central and satellite peaks on negative/positive $l$ were plotted as a function of temperature for both thermal processes in Fig.~\ref{fig:im}(d), 
where the setups A and B correspond to the results obtained on SENJU and WAND, respectively.
These data manifest the existence of clear hysteresis on the splitting of the magnetic peak. 
For example, the satellite peaks in the cooling process clearly survive down to 2.25~K.
On the other hand, only the central peak was observed up to 2.5~K upon heating from the base temperature.
The satellites emerge along with the central peak at 2.75~K, followed by the disappearance of the central peak at 3.0~K.
Despite a difference in the transition temperature, the magnitude of the splitting is very similar between both thermal processes.
The integrated intensity of the central and negative $l$ side of the satellite peak for the cooling process is displayed as a function of temperature in Fig.~\ref{fig:im}(e).
An abrupt evolution of the central peak and its coexistence with the satellite peak around 2.3~K are demonstrated in this behavior.
The existence of hysteresis and coexistence of central and satellite peaks confirm the first-order nature of the order-order transition at $T^*_{\rm N}{\sim}$2.5~K.
In contrast, no evident hysteresis is seen at {\tn}.

\begin{figure}[!ttt]
\includegraphics[width=8cm]{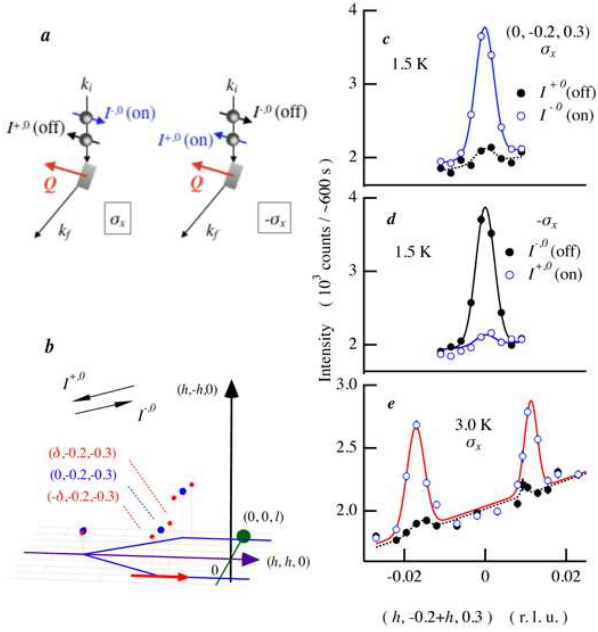}
\caption{(Color online) (a) Schematic drawing of the half polarized scattering setup at HB-1. Neutron spin is polarized parallel/antiparallel to the scattering vector {\qq}. Setup for -${\sigma}$ was realized by reversing the sign of the guide field.
(b) Illustrative scan trajectory and positions of magnetic peaks in the ground state (blue) and the intermediate phase (red) with respect to the horizontal ($h,~h,~l$) scattering plane.
(c-e) Incident polarization dependence of scans along ($h,~h,~0$) across (0,~-0.2,~-0.3) measured at (c)~1.5~K, (d)~1.5~K with reversed guide field, and (e)~3.0~K with original guide field. \vspace{-1cm}}
\label{fig:hp}
\end{figure}

Further insights into the nature of this additional transition in terms of magnetic structures were obtained by half-polarized neutron scattering experiments at HB-1.
The usual setup of the triple-axis spectrometer has limited access to the region beyond the horizontal scattering plane, whereas the magnetic peaks in EuPtSi spread into the three dimensional space due to the complex ordering vector.
In order to gain access to several magnetic and nuclear peaks, ($h,~h,~l$) was chosen as the horizontal scattering plane.
Owing to relatively large beam divergence and acceptance of the vertical direction, we could observe a series of out-of-plane magnetic reflections.
We focused on the nuclear and magnetic peaks at (2,~2,~0), (0,~0.2,~0.3), and (${\pm}{\delta}$, 0.2, 0.3), in which the magnetic peaks are placed slightly out-of-plane, as illustrated in Fig.~\ref{fig:hp}(b).
The incident polarization is fixed parallel or antiparallel to the in-plane component of {\qq}, which corresponds to the $x$ channel,  as displayed in Fig.~\ref{fig:hp}(a).
The polarization dependence of a scan along ($h,~h,~0$) through (0,~-0.2,~-0.3) at 1.5~K and 3.0~K are summarized in Fig.~\ref{fig:hp}(c-e).
$I_x^{{\pm},0}$ denotes the intensity of incident polarization + and - for channel $x$ without polarization analysis for the scattered beam, where the signs + and - correspond to the parallel and antiparallel directions, respectively. 
At 1.5~K in the ground state, a single magnetic peak was obtained for a polarization channel of $I^{-,0}$, which is the neutron spin antiparallel to the in-plane component of {\qq} obtained by activating the flipper.
This feature is further confirmed by reversing the polarization of the neutron guide field (-${\sigma}_x$), where the same state $I^{-,0}$ is realized by inactivating the flipper, as seen in Fig.~\ref{fig:hp}(a).
In these two contrasting setups, the magnetic peak at (0,~-0.2,~-0.3) is dominant for $I^{-,0}$, the neutron spin antiparallel to {\qq}.
The resulting average flipping ratio for this peak is 11${\pm}$2.
By heating up to 3.0~K in the intermediate phase, the magnetic peak was split into (${\pm}$0.03, -0.2, -0.3), which corresponds to (${\pm}h$,~-0.2${\pm}h$,~-0.3) with $h=0.015$ in the present geometry.
Both split peaks at 3.0~K have the same polarization dependence, as in the ground state.
The flipping ratio for the peaks at  (-0.03, -0.2, -0.3) and ( 0.03, -0.2, -0.3) are 10${\pm}$3 and 14${\pm}$3, respectively.
Therefore, the commensurate-incommensurate transition at {\tsn} does not affect the polarization dependence.

In order to reveal the nature of the field-induced $A$ phase in the phase diagram of Fig.~\ref{fig:phA}(a), 
single crystal neutron diffraction experiments were performed under vertical magnetic fields applied along [1,~1,~1]  on the renewed diffractometer, WAND$^2$.
Figure~\ref{fig:phA}(b) displays a part of the neutron scattering intensity map of the horizontal scattering plane defined by two orthogonal axes [1,~${\bar 1}$,~0] and [1,~1,~${\bar 2}$] recorded at 1.2~T, 1.9~K.
At zero field at 1.9~K, none of the magnetic peaks with {\qone}, which are illustrated as small balls in Fig.\ref{fig:phA}(c), are located in the horizontal scattering plane.
By ramping up to 1.2~T into the center of the $A$-phase, several magnetic peaks emerge in this scattering plane, as indicated by arrows in Fig.\ref{fig:phA}(b).
Around a nuclear peak at (1,~${\bar 1}$,~0), the superlattice peaks form hexagon, which can be indexed clockwise from top as $(0.79,~-0.71,~-0.08)$, $(1.08,~-0.79,~-0.29)$, $(1.29,~-1.09,~-0.20)$, 
$(1.20,~-1.28,~0.08)$, $(0.91,~-1.19,~0.28)$, and $(0.70,~-0.90,~0.20)$.
No further peak was observed both in-plane and out-of-plane around (1,~${\bar 1}$,~0). 
Similar hexagonal patterns around (1~0~${\bar 1}$), and four peaks out of six around (0,~0,~0) were observed as well.
This result indicates that magnetic peaks of the $A$-phase for {\bb}${\parallel}$[1,1,1] can be characterized  by the ordering vector {\qaeq}, in which only cyclical permutation is allowed along with retention of the order of the signs.
An important feature is that the active ordering vector is orthogonal to the applied magnetic field among equivalent vectors, illustrated as large balls in Fig.~\ref{fig:phA}(c). 

The possible magnetic structures in zero field are discussed henceforth.
The present polarized neutron scattering experiments reveal that the magnetic peak is dominated by the single polarization channel $I_x^{-,0}$ in both the ground state and the intermediate phase.
The scattering intensity of a helimagnet for a half-polarized setup for the $x$ channel is represented as
 \begin{equation}
 I_x^{{\pm},0}({\bm Q})~{\propto}~{\mu}_y^2 + {\mu}_z^2~{\pm}~2{\mu}_y{\mu}_z({\bm e_p} \cdot {\bm e_z}),
 \end{equation}
where ${\mu}_y$ and ${\mu}_z$ are the magnetic moments perpendicular to {\qq} in the horizontal plane and the out-of-plane, respectively.
The last term consists of the unit vectors of neutron polarization ${\bm e_p}$ and the vector chirality of the helical structure ${\bm e_z}$, in which the positive sign corresponds to a clock-wise helix with respect to the ordering vector {${\bm q}$}. 
The average flipping ratios for $I_x^{{\pm},0}$ of the magnetic peaks around (0,~-0.2,~-0.3) displayed in Fig.~\ref{fig:hp}(c-e) are 11${\pm}2$ and 12${\pm}4$ in the ground state and the intermediate phase, respectively.
These values are slightly lower than the flipping ratio of the nuclear 2 2 0 reflection of 18.4 measured by a Heusler analyzer.
Note that as shown in Fig.~\ref{fig:hp}, the magnetic peaks of both phases are located beyond the horizontal scattering plane of ($h,h,l$).
Therefore, the present polarization $x$ is not perfectly parallel to the scattering vector {\qq}.
By considering this offset of the magnetic peak, one can conclude that this magnetic peak is fully polarized for  $I^{-,0}$.
The result can be interpreted as ${{\mu}_y}{\simeq}-{{\mu}_z}$ with, single chirality with the counter-clockwise helix: $I_x^{-,0}{\propto}4{{\mu}_y}^2$ while $I_x^{+,0}=0$.
In other words, the magnetic moments lying within the plane normal to {\qone} have a helical structure with single counter-clockwise helicity.
This is consistent with the case of a single chirality in the crystal structure.
Representation analysis was performed to gain further information on the magnetic structure, but it does not impose additional constraints due to low symmetry of the present ordering vector.
In order to obtain conclusive magnetic structures, distinction between single- and multi-${\bm q}$ as well as crystal chirality is indispensable.

The present study revealed the long-range antiferromagnetic ground state with {\qoeq}. 
Although the lock-in transition at {\tsn} is clearly first order with hysteresis, no evident hysteresis was seen at {\tn}. 
An intensive search around {\qone} at 4.1~K just above {\tn} failed to detect any hint of partial order in a wide reciprocal space within the present accuracy.
It indicates that the remaining entropy above {\tn} does not arise from the short-range correlation around {\qone}, and its origin remains unclear.
On the other hand, the first-order lock-in transition from {\qop} to {\qone} is revealed, although no bulk measurements, such as specific heat and magnetization, detect this transition. 
This indicates that the entropy released at this transition {\tsn} can be small.
The ordering vectors {\qone} and {\qop} are different from that expected, assuming geometrical frustration\cite{Hopkinson2006,Isakov2008,Redpath2010},
and require further study to clarify their  origin.

In the case of MnSi, the ground state is a helimagnet with the propagation vector ${\bm q}$=$({\zeta},~{\zeta},~{\zeta})$ with ${\zeta}$=0.017, 
having a longer periodicity of ${\sim}$180~{\AA}\cite{Ishikawa1976}.
The half-polarized neutron scattering experiment proved that this helimagnetic structure has single chirality like the crystal structure\cite{Ishikawa1976,Shirane1983,Ishikawa1984,Ishida1985}.
By applying a magnetic field to enter into the $A$-phase, the magnetic peaks move into the plane normal to the applied field and form a hexagon based on a triple-${\bm q}$ structure, as revealed in small-angle neutron scattering experiments\cite{Muhlbauer2009,Nakajima2017}.
One characteristic of the magnetic peaks in the $A$ phase of MnSi is that the periodicity of the magnetic structure is unchanged when it moves into the plane perpendicular to the applied field.

The magnetic structure of EuPtSi in the ground state is revealed to be a helimagnet with {\qoeq}.
The ordering vector of {\qa} is different from that of MnSi; in particular, the periodicity is shorter by a factor of 10,
which appears to be a characteristic of helimagnets in $f$-electron systems, as observed in the chiral soliton lattice state in Yb(Ni$_{1-x}$Cu$_x$)$_3$Al$_9$\cite{Matsumura2017}. 
The presence of the intermediate phase in the zero field is also a unique feature of EuPtSi.
On the other hand, a helimagnetic structure with single chirality is common between two compounds.
In addition, the behavior upon entering the $A$-phase, namely, the magnetic ordering vector becomes orthogonal to the applied field while retaining the periodicity, is identical.
This fact, in addition to the emergence of additional Hall resistivity, implies that the $A$ phase in EuPtSi hosts the skyrmion lattice as well.
Further investigations, such as those on detailed magnetic field/temperature variations, are essential to unveil the nature of the $A$ phase in EuPtSi, and these are currently in progress.

\begin{acknowledgments}
We acknowledge T. Nakajima, K. Ohishi, T. U. Ito, T. Sakakibara, T. Takeuchi, Y. Homma, T. Matsumura, H. Nakao, C. Tabata, and R. Shiina for stimulating discussions. 
The authors thank the sample environment groups at HFIR, ORNL, and MLF J-PARC for their support on the neutron scattering experiments.
This work was partly supported by MEXT/JSPS KAKENHI Grant Numbers JP16K05031(Scientific Research (C)) and JP18H04329 (J-Physics), 
and used resources at the High Flux Isotope Reactor, a DOE Office of Science User Facility operated by the Oak Ridge National Laboratory.
WAND and its successor WAND$^2$ are operated jointly by ORNL and Japan Atomic Energy Agency under US-Japan Cooperative Program on Neutron Scattering. 
The neutron diffraction experiment on SENJU at MLF J-PARC was performed under a user program (Proposal No. 2018A0221).

\end{acknowledgments}

\bibliographystyle{./jpsj}

\end{document}